\documentclass[a4paper, conference]{IEEEtran}
\IEEEoverridecommandlockouts
\usepackage{cite}
\usepackage{graphicx}

\usepackage{amsmath}
\usepackage{mathtools}
\usepackage{algpseudocode}
\usepackage{algorithm}
\usepackage{algorithmicx}
\usepackage{textcomp}
\usepackage{filecontents}
\usepackage{microtype}
\usepackage{float}
\usepackage{adjustbox}
\usepackage{booktabs,makecell,tabularx}
\usepackage{url}
\usepackage{booktabs}
\usepackage{subfigure}

\newcolumntype{C}[1]{>{\centering\arraybackslash}p{#1}}
\newcolumntype{L}{>{\raggedright\arraybackslash}X}
\usepackage{siunitx}
\usepackage[utf8]{inputenc}
\renewcommand{\thetable}{\arabic{table}}
\usepackage{etoolbox}
\usepackage{xparse}
\makeatletter

    \makeatletter
    \newcommand{\linebreakand}{%
      \end{@IEEEauthorhalign}
      \hfill\mbox{}\par
      \mbox{}\hfill\begin{@IEEEauthorhalign}
    }
    \makeatother

\begin{document}




\title{Promoting Mental Well-Being for Audiences in a Live-Streaming Game by Highlight-Based Bullet Comments
    }
    \author{
\IEEEauthorblockN{Junjie H. Xu\IEEEauthorrefmark{1}\IEEEauthorrefmark{2}, Yulin Cai\IEEEauthorrefmark{1}\IEEEauthorrefmark{3}, Zhou Fang\IEEEauthorrefmark{2}, Pujana Paliyawan\IEEEauthorrefmark{4}}
\IEEEauthorblockA{
\IEEEauthorrefmark{2}Graduate School of Comprehensive Human Sciences, University of Tsukuba, Japan\\
\IEEEauthorrefmark{3}Graduate School of Information Science and Engineering, Ritsumeikan University, Japan\\
\IEEEauthorrefmark{4}Research Organization of Science and Technology, Ritsumeikan University, Japan\\
caiyulin.ritsumei@outlook.com}
}

    \maketitle
\begingroup\renewcommand\thefootnote{\IEEEauthorrefmark{1}}
\footnotetext{Equal contribution}
\endgroup

\begin{abstract}
This paper proposes a method for generating bullet comments for live-streaming games based on highlights (i.e., the exciting parts of video clips) extracted from the game content and evaluate the effect of mental health promotion. Game live streaming is becoming a popular theme for academic research. Compared to traditional online video sharing platforms, such as Youtube and Vimeo, video live streaming platform has the benefits of communicating with other viewers in real-time. In sports broadcasting, the commentator plays an essential role as mood maker by making matches more exciting. The enjoyment emerged while watching game live streaming also benefits the audience’s mental health. However, many e-sports live streaming channels do not have a commentator for entertaining viewers. Therefore, this paper presents a design of an AI commentator that can be embedded in live streaming games. To generate bullet comments for real-time game live streaming, the system employs highlight evaluation to detect the highlights, and generate the bullet comments. An experiment is conducted and the effectiveness of generated bullet comments in a live-streaming fighting game channel is evaluated.
\end{abstract}


\section{Introduction}


Game live streaming is one of the most popular categories in video live streaming. The spectators of game live streaming are increasing year by year \cite{Zorah}. Existing works have studied the positive relation between watching live streaming and mental health \cite{Mark}. Studies have shown that watching live-streaming video would reduce mental loneliness and create social connection \cite{William}. Based on these finding, watching live-streaming has a great potential in helping audiences to enhance mental health. Automatically commentary generation techniques applied in many activities. However, in terms of humor and audience experience, there are many rooms for improvement.

%

The bullet comment, a type of audience created commentary, is well accepted in live streaming platforms. The bullet comments are shooting from one side to the other side overlaid on top of the live streaming video. Because the bullet comments are attached to the video's timeline, the audience will be able to read others' commentary when live streaming is ongoing. It creates a shared viewing experience that benefits watching engagement \cite{Chen}. 

It is predictable that in the future of many games live-streaming channels, the actual player could be controlled by AI. Procedural Play Generation(PPG) aims to automatically generate entertaining game plays for the audiences \cite{Thawonmas}. The PPG is still under research area and it is considered important research area in live streaming. Currently, the gameplay generated by AI for live streaming lacks entertainment. As one possible solution, adding bullet comments to gameplay created by PPG could attract more viewers compared to gameplay only. 

The main contributions of this work are as follows:

\begin{enumerate}
\item This paper presents an automatic bullet comment generation system for game live streaming on the system generates the gameplay with skillful AI player combined with entertaining commentaries. 
\item Based on the current score of highlight and other game information, the system generates proper bullet comments for the audiences. We test our proposed system on a fighting video game in a live streaming platform.
\end{enumerate}

\begin{figure*}[htbp]
\centering
\begin{minipage}{8cm}
\centering
\includegraphics[width=1\textwidth]{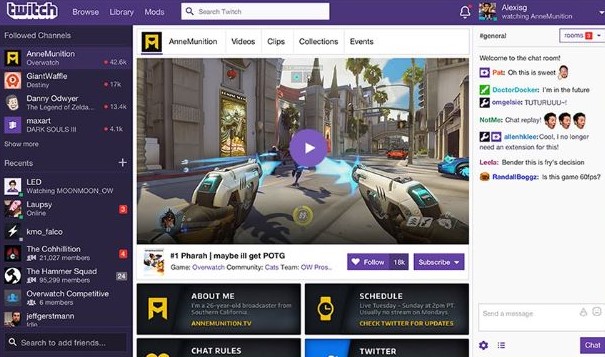}
\caption{Chat room style} \label{fig:twitch}
\end{minipage}
\begin{minipage}{9cm}
\centering
\includegraphics[width=1\textwidth]{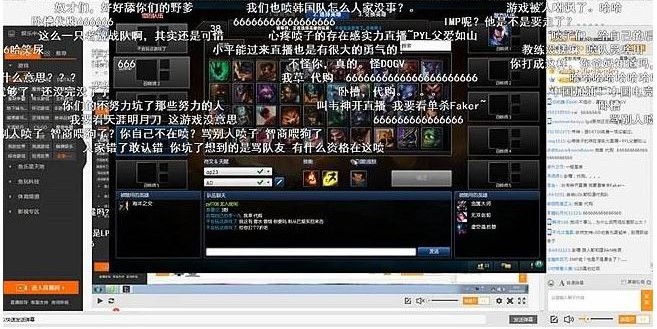}
\caption{Bullet comment style} \label{fig:danmu}
\end{minipage}
\end{figure*}

\section{Bullet Comments}
Generally, there are two forms of presentation of comments on live-streaming platforms. Chat room style and bullet comment style. Chat room style (Fig. \ref{fig:twitch}) is shown on the side of the video. Bullet comment style (Fig. \ref{fig:danmu}) is displayed at the top of the video and is called bullet-comment or Danmaku~\cite{wu2018}. Bullet comments are well-known on Chinese and Japanese live broadcast platforms such as Bilibili\textsuperscript{TM} and Niconico\textsuperscript{TM}. Different from the chat room styles such as Twitch\textsuperscript{TM} and Youtube\textsuperscript{TM}, the comments fly from the right to the left and are superimposed on the top of the video during the live broadcast, giving the audience a sense of realism-time interaction. These bullet comments usually consist of short sentences or just a few words, and in most cases, consist of slang and emotional expressions. Because bulleted comments are attached to the video’s timeline, viewers will be able to read other people’s comments on specific timeline points while the live broadcast is in progress. It creates a shared viewing experience that is conducive to viewing participation~\cite{Chen}.

Researchers studied why bullet comments are famous in China and Japan~\cite{yang2019}. The information entropy of Chinese characters is higher than the English vocabulary ~\cite{sun2019compound}. Because the bullet comment period in the video is short, Chinese characters can load more information than English. If the sentence structure and complexity are similar, there are more keywords in English than in Chinese. When viewers read bullet comments, they mostly read them quickly. Extracting keywords can intuitively understand the meaning of comments. When the audience reads the English comments, although the keywords can be extracted, the association is complicated, and the audience cannot immediately understand what these English comments are saying.

\section{Proposed System}

The fundamental idea is to apply the predefined rules to pattern match the comment templates with the game states. If the game situations match the predefined templates, the system will generate bullet comments and filling up key positions game scene. Also, the system will calculate the highlight score of the game in real-time. If the current highlight score passes the threshold of the highlight level, the system will generate the highlight bullet comments.

\subsection{Bullet Comment collection:} 

Professional commentaries are collected from the online videos of SFV (Street Fighter V)  which is one of the famous fighting game. The recorded video is available on Youtube with auto-transcript commentaries.  During the game match, the commentators will give essential commentaries regarding the situation about both of the players. The move descriptions from these commentaries were downloaded and filtered manually. 
We utilize an open data set collected from Bilibili (Chinese video-sharing website with bullet comments) introduced by Qing et al. \cite{Qing} as our highlight bullet comments. It contained emotional words often used in bullet comments. The positive emotion words combined with the above mentioned templates to build the bullet comment bank for generating text.

\subsection{Highlight detection:} 

\begin{figure}[htbp]
\centering
\begin{minipage}{4cm}
\centering
\includegraphics[width=1\textwidth]{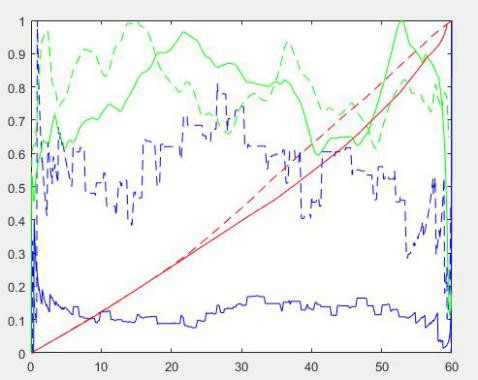}
\caption{Six Features Cues} \label{fig2}
\end{minipage}
\begin{minipage}{4cm}
\centering
\includegraphics[width=1\textwidth]{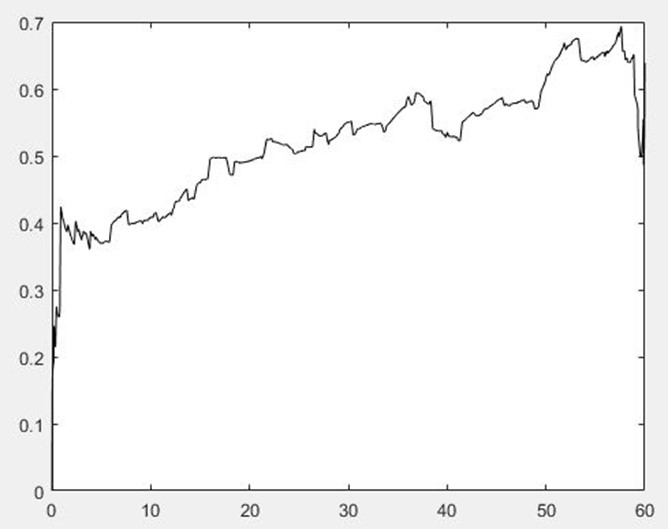}
\caption{Combined Highlight Cues} \label{fig3}
\end{minipage}
\end{figure}

Adapted from Chu et al.\cite{Chu}. There are three important factors that affect the highlight level in a round of fighting game. Game--action, distance, and Hit-Point(HP). Because the game consists of two players, a total of six features are extracted. The estimation of the current highlight score is followed by the previous works~\cite{Ishii}. The straight line is player one’s cues, while the dashed line is player two’s cues. The average of highlight cues variation derived from 883 rounds of a game round and curve of the combined highlight value is shown in Fig. 1-2. The color of the score cue, distance cue, and game--action cue are red, green, and blue respectively (See Fig. \ref{fig3}). We set the threshold value for the highlight detection based on the pre-conducted simulation of the game. Adapting the model, we combined above mentioned fighting game factors to estimate the highlight of the fighting game. A total of six highlight cues is selected.
\begin{itemize}
    \item P1's score: The score cue of Player one.
    \item P2's score: The score cue of Player two.
    \item P1's Action: The action cue of Player one.
    \item P2's Action: The action cue of Player two.
    \item P1's Distance: The distance cue of Player one.
    \item P2's Distance: The distance cue of Player two.
\end{itemize}
These cues describe the level of the highlight from a different perspective. We can jointly consider all perspectives by combining them. The highlight score of the \textit{k} time in the game can be calculated as follows:
\begin{equation}
    H(k)=F(\hat{C_i(k)}),i=1,...,6;k=1,...N
\end{equation}
The \^{C}\textsubscript{i}(k)is the \textit{i} cue value \textit{C}\textsubscript{i} after the smooth process, the function \textit{F} is for combining highlight cues from various perspectives. The smooth process is defined as follows:
\begin{equation}
    \hat{C_i(k)} =\frac{\max_k(C_i(k))}{\max_k(C_i(k)^{CMA})} \times C_i(k)^{CMA}
\end{equation}
\begin{equation}
    C_i(k)^{CMA} = \frac{C_i(1)+...+C_i(k)}{k},k=1,...,N
\end{equation}
The \textit{CMA} is for the cue averaging. By applying this, the change of highlight value will not be abruptly for consecutive gameplay. The normalization defined in (Eqn. 3) makes different highlight cues integrate easily. The function \textit{F} is a simple linear combination and defined as follows:
\begin{equation}
    F(\hat{C_i(k)}) =\frac{1}{6}\sum_{i=1}^{6} \hat{C_i(k)}
\end{equation}

\section{User Evaluation and Results}

\begin{figure}[htbp]
\centering
\begin{minipage}{4cm}
\centering
\includegraphics[width=1\textwidth]{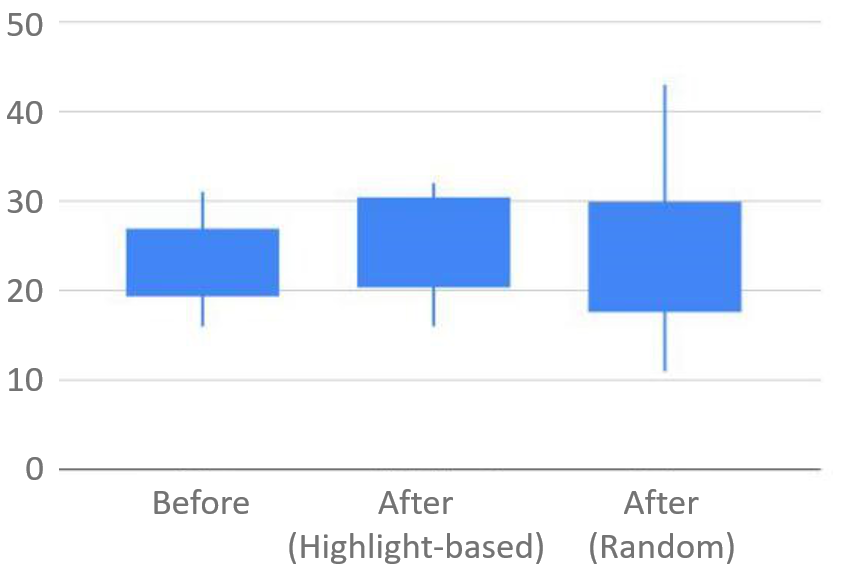}
\caption{Negative affect of PANAS} \label{fig4}
\end{minipage}
\begin{minipage}{4cm}
\centering
\includegraphics[width=1\textwidth]{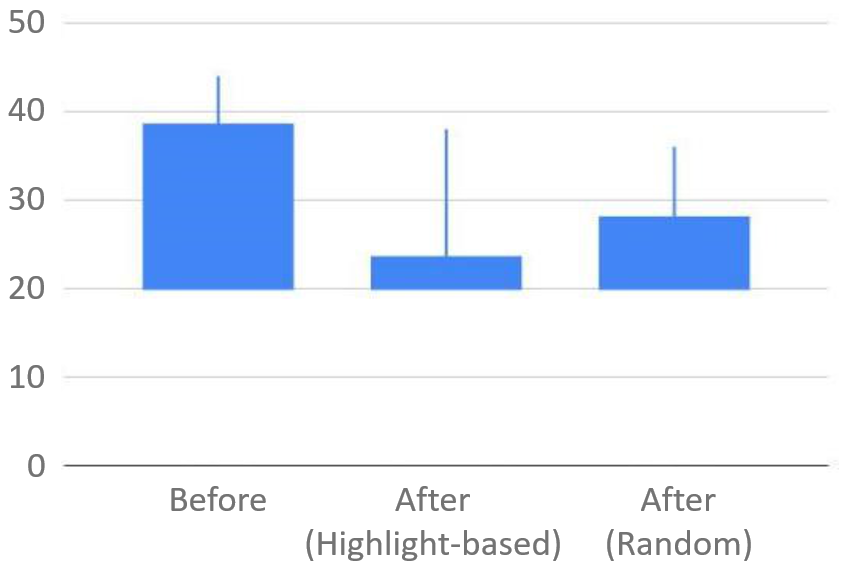}
\caption{Positive affect of PANAS} \label{fig5}
\end{minipage}
\end{figure}

An user evaluation was conducted by the positive and negative affect schedule (PANAS) questionnaire. PANAS is a self-report measure of positive and negative affect made by Watson et al. \cite{PANAS}. It consists of 10 positive emotions and 10 negative emotions, and each of them was asked in 5 points Liker-type scale (1 = Very Slightly or Not at all, 5 = Extremely). The FightingICE, a fighting game platform used in fighting game competition, was used and live-streaming in Twitch. In this experiment,  Chinese bullet comments were used for the pilot test and a group of eight healthy graduate students participated the experiment. The proposed system was compared with the random based bullet comment generation. 
The result of before and after watching the live-streaming is shown on Fig. 3-4.  The reliability of whole PANAS questions was evaluated by using Cronbach's alpha~\cite{Cronbach}. The average reliability was 0.821, which is good. 
Before watching any live-streaming video, the result of the average score of the positive affect is 22.87, and the negative affect is 14.625. After watching the highlight-based live-streaming, the result of the average score of the positive affect is 24.625 and negative affect is 11.875. As for the random bullet comments live-streaming, the result is 25.5 for positive and 12.375 for negative. 
From the results, watching highlight-based live-streaming can reduce the negative affect better than the live-streaming without applying highlight-based generation. As for the positive affects, there is no much difference between these two approaches.  The lower the negative affect, the better the result. From this viewpoint,  watching game live-streaming with highlight-based bullet comments can promote mental health in a certain degree.

\section{Conclusion and Future Work}
The research proposed an highlight-based bullet comment generation for live-streaming game to promote the mental health of spectators. The system was test and live streaming it on Twitch. For comparison, two types of bullet comments generation policies were tested. From the results of PANAS, the highlight-based bullet comment generation performed better than the random bullet comment generation in terms of reducing negative affects. 

There are some limitations to this work. Highlight-based bullet comment generation was tested in a fighting game, and other types of games also have the potential to combine with highlight-based bullet comments. Other language bullet comments can be evaluated the effects. A more general system design that suited for other languages is the future works need to focus on.


\begin{thebibliography}{99}

\bibitem{Zorah} Hilvert-Bruce, Z., Neill, J. T., Sjöblom, M., and Hamari, J, ``{\it Social motivations of live-streaming viewer engagement on Twitch},'' In {\it Computers in Human Behavior}, 84, 58-67, 2018.

\bibitem{Mark} Johnson, M. R, ``{\it Inclusion and exclusion in the digital economy: Disability and mental health as a live streamer on Twitch. tv},'' In {\it Information, Communication and Society}, 22(4), 506-520, 2019.


\bibitem{William} Hamilton, W. A., Garretson, O., and Kerne, A, ``{\it Streaming on twitch: fostering participatory communities of play within live mixed media},'' In {\it Proc. CHI}, pp. 1315-1324, 2014.



\bibitem{Chen} Chen, Y., Gao, Q., and Rau, P. L. P, ``{\it Understanding gratifications of watching danmaku videos–videos with overlaid comments},'' In {\it Proc. International Conference on Cross-Cultural Design}, pp. 153-163, 2015.


\bibitem{Thawonmas} Thawonmas, R., and Harada, T, ``{\it AI for Game Spectators: Rise of PPG},'' In {\it Proc. Workshops at the Thirty-First AAAI Conference on Artificial Intelligence}, pp. 1032-1033, 2017.

\bibitem{wu2018} Wu, Q., Sang, Y., Zhang, S., and Huang, Y, ``{\it Danmaku vs. forum comments: understanding user participation and knowledge sharing in online videos},'' In {\it Proceedings of the 2018 ACM conference on supporting groupwork}, pp. 209-218, 2018.

\bibitem{yang2019} Yang, Y, ``{\it Danmaku subtitling: An exploratory study of a new grassroots translation practice on Chinese video-sharing websites},'' {\it  Translation Studies, 14(1)}, pp. 1-17, 2021.

\bibitem{sun2019compound} Sun, C. C., and Hendrix, P, ``{\it Compound words in Mandarin Chinese and English: the role of information theory 6th Nov.},'' In {\it Book of Abstracts}, pp. 55, 2019.

\bibitem{Qing} Ping, Q., and Chen, C, ``{\it Video highlights detection and summarization with lag-calibration based on concept-emotion mapping of crowd-sourced time-sync comments},'' {\it Proc. Workshop on New Frontiers in Summarization}, pp. 1-11, 2017.


\bibitem{Chu} Chu, W. T., and Chou, Y. C, ``{\it On broadcasted game video analysis: event detection, highlight detection, and highlight forecast.},'' In {\it Multimedia Tools and Applications, 76(7)}, pp. 9735-9758, 2017.

\bibitem{Ishii} Ishii, R., Ito, S., Thawonmas, R., and Harada, T, ``{\it A Fighting Game AI Using Highlight Cues for Generation of Entertaining Gameplay},'' In {\it Proc. IEEE Conference on Games (CoG)}, pp. 1-6, 2019.



\bibitem{PANAS} Watson, D., Clark, L. A., and Tellegen, A, ``{\it Development and validation of brief measures of positive and negative affect: the PANAS scales},'' In {\it Journal of personality and social psychology}, 54(6), 1063, 1988.

\bibitem{Cronbach} Cronbach, L. J, ``{\it Coefficient alpha and the internal structure of tests},'' In {\it psychometrika}, 16(3), 297-334, 1951.


\end{thebibliography}
\end{document}